\documentstyle[aps,epsfig,amsfonts,epsf,feynmf]{revtex}

\def\cm#1{}
\newcommand{\p}{\partial}
\newcommand{\f}[2]{\frac{#1}{#2}}
\newcommand{\be}{\begin{eqnarray}}
\newcommand{\ee}{\end{eqnarray}}
\newcommand{\beqn}{\begin{eqnarray}}
\newcommand{\eeqn}{\end{eqnarray}}

\newcommand{\Zm}{Z_{m^2}}
\newcommand{\Zg}{Z_{g}}
\newcommand{\Zphi}{Z_{\phi}}
\newcommand{\Ze}{Z_{e}}
\newcommand{\Za}{Z_{A}}
\newcommand{\Zv}{Z_{v}}

\newcommand{\calL}{{\cal L}}
\newcommand{\Amu}{A_{\mu}}
\newcommand{\Anu}{A_{\nu}}
\newcommand{\phicross}{\phi^{\dagger}}
\newcommand{\PmunuT}{P_{\mu\nu}^T}
\newcommand{\Dmunu}{D_{\mu\nu}}
\newcommand{\mg}{m_{\gamma}}
\newcommand{\Tr}{\mbox{Tr}}
\newcommand{\ep}{\varepsilon}
\newcommand{\barln}{\overline{\ln}}
\newcommand{\muep}{\mu^{\ep}}
\newcommand{\db}{\,\,{\bar {}\!\!d}\!\,\hspace{0.5pt}}
\newcommand{\dbarp}{\db p\,}
\newcommand{\dbarq}{\db q\,}
\title{Two-Loop Effective Potential of O($N$)-Symmetric
Scalar QED in $4-\ep$
 Dimensions}

\author{H. Kleinert}
\address{Institut f\"ur Theoretische Physik,
Arnimallee 14 D-14195 Berlin,
 Germany}
\author{B. Van den Bossche}
\address{Physique Nucl\'eaire Th\'eorique, B5,
Universit\'e de Li\`ege Sart-Tilman,
 4000 Li\`ege, Belgium\\
and\\
Institut f\"ur Theoretische Physik,
Arnimallee 14 D-14195 Berlin,
 Germany}

\begin{document}

\maketitle

\begin{abstract}
The effective potential of scalar QED is computed analytically up to
two loops in the Landau gauge. The result is given
in $4-\ep$ dimensions using
minimal subtraction and $\ep$-expansions.
In three dimensions ($\ep=1$), our calculation is intended
to help throw light on unsolved problems
of the superconducting phase transition,
where critical exponents
and the position of the tricritical point have not yet
found a satisfactory explanation within the renormalization group
approach.
\end{abstract}

\section{Introduction}

Recently, important progress has been made to the evaluation of
Feynman diagrams in $4-\ep$ dimensions.
In particular, two-loop Feynman diagrams
with unequal masses of the internal lines are now available
analytically
\cite{jones92,tausk92}.
Moreover,
the full $\ep$-expansion of the sunset diagram is  known
\cite{davydychev00}.
 This is important for critical phenomena in $4-\ep$
dimensions since,
for instance, a three-loop calculation requires
the knowledge
of all terms of order
$\ep$ of the two-loop diagrams.

With the help of these results, we investigate in the present paper
an O($N$)-symmetric version of scalar QED containing $N/2$ complex fields
in $4-\ep$
 dimensions,
which, for $N=2$, reduces to the Ginzburg--Landau model
of superconductivity.
Our results are intended to shed
some light
on
the open problems
of the
superconducting phase transition in three dimensions.
{}From a dual formulation of the
Ginzburg-Landau theory we know
that there must exist a regime
of the parameter $ \kappa $, the ratio between
 scalar and vector masses, where the transition changes from
first to second order \cite{tri}.
This scenario
has
been
confirmed by several Monte-Carlo simulations, see
\cite{barto83}.
%\cite{olsson98}. and references therein.

In contrast to this, the
 traditional $4-\ep$ -dimensional
renormalization group analysis of the
Ginzburg-Landau model has produced
more  puzzles than answers.
At the one-loop order, a second-order transition
 is obtained only  for
 $N\ge N_c=366$, much larger than the $N=2$ -value of
a  superconductor
\cite{halperin74,lawrie82}.
The situation does not improve by going to two loops \cite{tessman84}.
Only a somewhat contrived resummation procedure leads to
a desired fixed point \cite{folk96}.
Only in
direct three-dimensional calculations has
 a second-order transition  been obtained
 for $N=2$, see for example
\cite{nog99}.
The absence of a charged fixed
point for $N=2$ in  $4-\ep$ dimensions
seems thus to be  a specific weakness
of this
approach.

At present it is hoped that variational
perturbation theory may be able to
locate a  fixed point
of the Ginzburg-Landau model, thus allowing to
extract physical values independently of Ref. \cite{folk96}.
This theory developed in Ref.~\cite{kleinertpi}
has proven to be a powerful tool for determining
 critical exponents  in
 three  \cite{kleinert257,kleinert279,kleinert287}
as well as in $4-\ep$ dimensions~\cite{kleinert263,kleinert295}.
Recently, we have applied it to the determination of amplitude ratios
in three dimensions of the O($N$)-model~\cite{kleinertvdb01}.
We did these calculations
using a
method proposed
by the Aachen group \cite{dohm99},
where analytic renormalization
 is applied
in the form of minimal subtraction
although working directly in  $D=3$ dimensions,
without any
$\ep$-expansion.
A similar
 determination of amplitude ratios
in  $4-\ep$ dimensions with $\ep$-expansion
is still missing.
For determining the associated amplitude ratios,
we need the effective potential in $4-\ep$ dimensions.
This is what we want to calculate here.
%Technically speaking, our paper is a direct extension
%of the one-loop work of Lawrie
%\cite{lawrie82}.

As a cross check we shall
often set the electric charge
equal to zero
and check that we recover
 the results for the  usual O($N$)-symmetric $\phi^4$-theory.

%It will however be necessary to give
%also the fixed points, i.e. to compute the beta functions (this is necessary
%also to achieve the first goal). Having the fixed points and the
%equation of state, which is readily obtained from the effective potential,
%we should be able answer the third project.

The renormalization constants and various beta functions of scalar QED have
 been calculated up to two loops
in the early work on this subject \cite{tessman84,folk92}.
It is not our purpose
to rederive these results.
However, in the process of obtaining the effective potential, i.e., of
calculating the various Feynman diagrams in the symmetry-broken phase,
we obtain not only the effective potential but also the
renormalization constants of
mass $\Zm$
and $\phi^4$-coupling constant $\Zg$ to
up to two loops.
The
renormalization constants of the scalar and vector fields
 $\Zphi$ and $\Za$, on the other hand,  enters  only
with the first loop order.
The fact that the charge coupling renormalization constant $\Ze$
is not obtained to the same
order as the $\phi^4$-coupling constant one has its origin
in  the fact that,
due to a Ward identity,  $\Ze=\Zphi$, see Eq.~(\ref{EQeren}).

Before closing the introduction, we point out that the effective potential
of the electroweak part of the standard model has already been
calculated to the two-loop order \cite{jones92}. However, our work is
more than the abelian $U(1)$ subset of this reference for two reasons: 
first, we work with $N/2$-complex scalar fields (N=2 in \cite{jones92});
second, in view of the application to critical phenomena in three dimensions,
the $\ep$-expansion is used. At the two-loop order, this means that the
one-loop part of the effective potential is expanded up to the
order $\ep$. This term is of course not there when 
being interested in going in four dimensions, as is the case of 
Ref.~\cite{jones92}.

Finally, we mention that our work can be seen as the extension
to a two coupling constants problem of the work of Br\'ezin et al. 
\cite{brezin72} (see also \cite{zj76}), which consider the $\ep$-expansion
of the equation of state of the $N$-components $\phi^4$ 
theory to the two-loop 
order\footnote{For the Ising model, corresponding to $N=1$, the authors
of \cite{zia73} have succeeded going to the three-loop order of the
$\ep$-expansion of the equation of state.}.
Without the gauge field, the effective potential can be used to determine
this equation of state.

The paper is organised as
follow: in Section~\ref{model}, we specify the model
and our conventions. In Section~\ref{calculation}, we present the various
intermediate steps for the obtaining of the effective potential.
The individual results are combined
in Section~\ref{effpot}, and the conclusions drawn in
Section~\ref{conclusion}.

\section{Model}
\label{model}
The Lagrangian density to be studied contains $N/2$ complex scalar fields
$\phi$
coupled to the abelian
fields $\Amu$ and reads, with a covariant gauge fixing,
\be
\calL=|D\phi|^2+m^2\phi^2+\f{g}{3!}|\phi|^4+\f{1}{4}F_{\mu \nu
}^2+\f{1}{2\alpha}
(\p_\mu A_\mu)^2,
\label{EQqedlag}
\ee
where $D_{\mu}=\p_{\mu}-ieA_{\mu}$ is the covariant derivative, $F_{\mu \nu }$
is the
usual field-strength tensor and $\alpha$ is a gauge parameter.
There is no need to
introduce  ghost fields
to guarantee  gauge invariance since these decouple from the system
in  an abelian gauge  theory.

The effective potential will be obtained using the so-called
{\em background-field method\/}
of DeWitt \cite{DeWitt}. We shift the scalar
field by an unknown constant $\Phi$:
$\phi\rightarrow \Phi+\phi$. This generates new vertices. To simplify the
calculation, we shall use throughout the Landau gauge $\alpha\rightarrow 0$.
This reduces the number of Feynman diagrams
and,
since $ \alpha =0$ enforces
$\p_\mu A_\mu\equiv 0$ at the  Lagrangian level,
removes a possible mixing  of
$\Amu\phi$ and $\Amu\phicross$ terms,
thus decoupling
scalar and gauge propagators. It is further
advantageous to use real fields, defining
\beqn
\phi=\f{1}{\sqrt{2}}(\phi_1+i\phi_2),
{}~~~~~~~~~~
\Phi=\f{1}{\sqrt{2}}(\Phi_1+i\Phi_2).
\eeqn
Then the Lagrangian has the expansion
around the background field:
\beqn
\calL&=&\calL_0+\f{1}{2}\phi[G^TP^T+G^LP^L]\phi+
\f{1}{2}\Amu D^T\PmunuT \Anu\nonumber\\
&&\mbox{}+\f{g}{4!}[(\phi^2)^2+4\Phi\phi(\phi^2)]
+e^2A^2\Phi\phi
+\f{1}{2}e^2A_\mu^2\phi^2+eA_{\mu}(\phi_2\p_{\mu}\phi_1-\phi_1\p_{\mu}\phi_2),
\label{lagrangian}
\eeqn
where $\phi$ and $\Phi$ are now $N$ components real fields
written as two-dimensional iso-vectors
 $\phi=(\phi_1,\phi_2)$ and $\Phi=(\Phi_1,\Phi_2)$. The
notation is:
\beqn
\calL_0&=&\f{1}{2}m^2\Phi^2+\f{g}{4!}\Phi^4\\
G^T&\equiv&-\p^2+m_T^2=-\p^2+m^2+\f{g}{3!}\Phi^2,
\label{EQdefmT}~~~~
G^L\equiv-\p^2+m_L^2=-\p^2+m^2+\f{g}{2}\Phi^2,
\label{EQdefmL}\\
D^T&\equiv&-\p^2+\mg^2=-\p^2+e^2\Phi^2,
\label{EQdefmg}\\
P^T_{ij}&=&\delta_{ij}-\f{\Phi_i\Phi_j}{\Phi^2},~~~~
P^L_{ij}=\f{\Phi_i\Phi_j}{\Phi^2},~~~~
\PmunuT=\delta_{\mu\nu}-\f{\p_{\mu}\p_{\nu}}{\p^2},
\eeqn
with $G^T,G^L$ being
the transverse and longitudinal inverse propagators of
the scalar field, and $\Dmunu^T$ the inverse transverse propagator of the
photon
field. The transversality of the latter is due to  the Landau gauge.
Note that there is no  term $e\Amu\p_{\mu}(\Phi_2\phi_1-\Phi_1\phi_2)$
which would mix
vector and scalar propagators.

Compared to the complex-field representation, real
fields have one small complication:
the gauge-scalar-scalar vertex
 involves a vector product since it mixes the real and imaginary parts $\phi_1$
and $\phi_2$.
For complex fields
 it is diagonal in
the scalar field space.

Up to two loops, the fastest way
of determining the Feynman diagrams
with their proper weight is to use the Wick theorem. In higher loops,
it might be interesting to use a
more efficient algorithm to generate the diagrams and their
weight, in the same spirit as, for instance, \cite{ka00,ka00b}.
Using a similar
procedure, we have derived the diagrams of scalar QED up to the fourth order.
Their calculation will form the subject of a separate publication
\cite{vdb01b}.
%The connected one-particle reducible diagrams are not considered since we
%are interested in the effective potential.

We shall calculate the diagrams
in the following expansion,  in which
 $\hbar$ counts the number of loops:
%

%\ieps{1}{2}{pot}
\begin{fmffile}{graph1}
\fmfset{wiggly_len}{2mm}
\fmfset{thin}{0.7pt}
\fmfset{dot_size}{1.0thick}
\beqn
V_{\mbox{eff. pot.}}&=&V_{\mbox{classical}}+\f{\hbar}{2}\biggl[(N-1)
\parbox{9mm}{\begin{center}
\begin{fmfgraph*}(18.5,18.5)
\fmfleft{i1}
\fmfright{o1}
\fmf{plain,left=1,label={\footnotesize $T$},label.dist=1mm}{i1,o1}
\fmf{plain,left=1}{o1,i1}
\end{fmfgraph*}\end{center}}
+
\parbox{9mm}{\begin{center}
\begin{fmfgraph*}(18.5,18.5)
\fmfleft{i1}
\fmfright{o1}
\fmf{plain,left=1,label={\footnotesize $L$},label.dist=1mm}{i1,o1}
\fmf{plain,left=1}{o1,i1}
\end{fmfgraph*}\end{center}}
+(D-1)
\parbox{9mm}{\begin{center}
\begin{fmfgraph*}(18.5,18.5)
\fmfleft{i1}
\fmfright{o1}
\fmf{photon,left=1}{i1,o1}
\fmf{photon,left=1}{o1,i1}
\end{fmfgraph*}\end{center}}
\biggr]
\nonumber\\
&&\hspace{-2cm}\mbox{}\hbar^2\Biggl\{
\f{g}{4!}\biggl[(N^2-1)
\parbox{15mm}{\begin{center}
\begin{fmfgraph*}(35,25)
\fmfleft{i1}
\fmfright{o1}
\fmf{plain,left=1,label={\footnotesize $T$},label.dist=1mm}{i1,v1}
\fmf{plain,left=1}{v1,i1}
\fmf{plain,left=1}{o1,v1}
\fmf{plain,left=1,label={\footnotesize $T$},label.dist=1mm}{v1,o1}
\fmfdot{v1}
\end{fmfgraph*}\end{center}}
+2(N-1)
\parbox{15mm}{\begin{center}
\begin{fmfgraph*}(35,25)
\fmfleft{i1}
\fmfright{o1}
\fmf{plain,left=1,label={\footnotesize $T$},label.dist=1mm}{i1,v1}
\fmf{plain,left=1}{v1,i1}
\fmf{plain,left=1}{o1,v1}
\fmf{plain,left=1,label={\footnotesize $L$},label.dist=1mm}{v1,o1}
\fmfdot{v1}
\end{fmfgraph*}\end{center}}
+3
\parbox{15mm}{\begin{center}
\begin{fmfgraph*}(35,25)
\fmfleft{i1}
\fmfright{o1}
\fmf{plain,left=1,label={\footnotesize $L$},label.dist=1mm}{i1,v1}
\fmf{plain,left=1}{v1,i1}
\fmf{plain,left=1}{o1,v1}
\fmf{plain,left=1,label={\footnotesize $L$},label.dist=1mm}{v1,o1}
\fmfdot{v1}
\end{fmfgraph*}\end{center}}
\biggr]
+\f{e^2}{2}(D-1)\biggl[
(N-1)
\parbox{15mm}{\begin{center}
\begin{fmfgraph*}(35,25)
\fmfleft{i1}
\fmfright{o1}
\fmf{plain,left=1,label={\footnotesize $T$},label.dist=1mm}{i1,v1}
\fmf{plain,left=1}{v1,i1}
\fmf{photon,left=1}{o1,v1,o1}
\fmfdot{v1}
\end{fmfgraph*}\end{center}}
+
\parbox{15mm}{\begin{center}
\begin{fmfgraph*}(35,25)
\fmfleft{i1}
\fmfright{o1}
\fmf{plain,left=1,label={\footnotesize $L$},label.dist=1mm}{i1,v1}
\fmf{plain,left=1}{v1,i1}
\fmf{photon,left=1}{o1,v1,o1}
\fmfdot{v1}
\end{fmfgraph*}\end{center}}
\biggr]\nonumber\\
&&\hspace{-1.8cm}\mbox{}
-\left(\f{g}{3!}\right)^2
\biggl[(N-1)
\parbox{14mm}{\begin{center}
\begin{fmfgraph*}(25,25)
\fmfleft{v1}
\fmfright{v2}
\fmf{plain,label={\footnotesize $L$},label.dist=0.0mm}{v1,v2}
\fmf{plain,left=0.7,label={\footnotesize $T$},label.dist=1mm}{v2,v1}
\fmf{plain,left=0.7,label={\footnotesize $T$},label.dist=1mm}{v1,v2}
\fmfdot{v1,v2}
\end{fmfgraph*}
\end{center}}
+3
\parbox{14mm}{\begin{center}
\begin{fmfgraph*}(25,25)
\fmfleft{v1}
\fmfright{v2}
\fmf{plain,label={\footnotesize $L$},label.dist=0.0mm}{v1,v2}
\fmf{plain,left=.7,label={\footnotesize $L$},label.dist=1mm}{v2,v1}
\fmf{plain,left=.7,label={\footnotesize $L$},label.dist=1mm}{v1,v2}
\fmfdot{v1,v2}
\end{fmfgraph*}
\end{center}}
\biggr]
-4\left(\f{e^2}{2}\right)^2\Phi^2
\parbox{14mm}{\begin{center}
\begin{fmfgraph*}(25,25)
\fmfleft{v1}
\fmfright{v2}
\fmf{plain,label={\footnotesize $L$},label.dist=0.0mm}{v1,v2}
\fmf{photon,left=.7}{v2,v1}
\fmf{photon,left=.7}{v1,v2}
\fmfdot{v1,v2}
\end{fmfgraph*}
\end{center}}
-e^2\biggl[
(N-2)\parbox{14mm}{\begin{center}
\begin{fmfgraph*}(25,25)
\fmfleft{v1}
\fmfright{v2}
\fmf{photon}{v1,v2}
\fmf{plain,left=.7,label={\footnotesize $T$},label.dist=1mm}{v2,v1}
\fmf{plain,left=.7,label={\footnotesize $T$},label.dist=1mm}{v1,v2}
\fmfdot{v1,v2}
\end{fmfgraph*}
\end{center}}
+2
\parbox{14mm}{\begin{center}
\begin{fmfgraph*}(25,25)
\fmfleft{v1}
\fmfright{v2}
\fmf{photon}{v1,v2}
\fmf{plain,left=.7,label={\footnotesize $T$},label.dist=1mm}{v2,v1}
\fmf{plain,left=.7,label={\footnotesize $L$},label.dist=1mm}{v1,v2}
\fmfdot{v1,v2}
\end{fmfgraph*}
\end{center}}
\biggr]
\Biggr\}.
\label{EQdiagrams}
\eeqn
\end{fmffile}

Explicitly, this equation reads in a unit $D$-dimensional volume:

\beqn
V&=&\f{1}{2}m^2\Phi^2+\f{g}{4!}\Phi^4
+\f{\hbar}{2}\left[(N-1)\Tr\ln(G^T)
+\Tr\ln(G^L)+(D-1)\Tr\ln(D^T)\right]\nonumber\\
&&\mbox{}+\hbar^2\Biggl(\f{g}{4!}
\left\{
\left[
(N-1)\int \dbarp\Delta(p,m_T)+\int \dbarp \Delta(p,m_L)
\right]^2+2
(N-1)\left[\int \dbarp\Delta(p,m_T)\right]^2
+2\left[\int \dbarp \Delta(p,m_L)
\right]^2
\right\}\nonumber\\
&&\mbox{}+\f{e^2}{2}(D-1)\left[\int \dbarp\Delta(p,\mg)\right]\left[
(N-1)\int \dbarp\Delta(p,m_T)+\int \dbarp\Delta(p,m_L)
\right]\nonumber\\
&&\mbox{}-\left(\f{g}{3!}\right)^2\Phi^2
\left[
(N-1)\int \dbarp\dbarq\Delta(p,m_T)\Delta(q,m_T)\Delta(p+q,m_L)
+3\int \dbarp\dbarq\Delta(p,m_L)\Delta(q,m_L)\Delta(p+q,m_L)
\right]\nonumber\\
&&\mbox{}-4\left(\f{e^2}{2}\right)^2\Phi^2
\left[
\int \dbarp \dbarq \PmunuT(p)\PmunuT(q)
\Delta(p,\mg)\Delta(q,\mg)\Delta(p+q,m_L)
\right]\nonumber\\
&&\mbox{}-e^2\left\{
\int \dbarp \dbarq \PmunuT(p)q_{\mu}q_{\nu}\Delta(p,\mg)\left[
(N-2)\Delta(q,m_T)\Delta(p+q,m_T)+2\Delta(q,m_L)\Delta(p+q,m_T)
\right]
\right\}\Biggr),
\label{EQalgebraic}
\eeqn
where $\dbarp\equiv d^Dp/(2\pi)^D$, and
$\Delta(p,m)=1/(p^2+m^2)$.

All  quantities in these expressions (fields, coupling constants, and masses)
are bare quantities.
Up to the second order in the loop expansion, the divergences show up
as poles
in $\ep$ up to the order $1/\ep^2$. They have to be removed
to have a finite limit $\ep\rightarrow 0$.
They may either be removed
by adding counter-terms  to the initial Lagrangian,
and calculating the associated Feynman diagrams
in addition to
those in (\ref{EQdiagrams}).
Alternatively, and this is how we shall proceed,
we may use~(\ref{EQdiagrams})
and~(\ref{EQalgebraic}) and include
renormalization constants
 which are reexpanded up to any
given order in the loop parameter, i.e., up to
the order $\hbar^2$ in our case.
These renormalization constants are defined as
\beqn
\phi=\phi_r\sqrt{\Zphi},~~~
\Amu={\Amu}_r\sqrt{\Za},~~~
m^2=m^2_r\f{\Zm}{\Zphi},~~~
g=g_r\mu^{\ep}\f{\Zg}{\Zphi^2},~~~
e=e_r\mu^{\ep/2}\f{\Ze}{\Zphi\sqrt{\Za}}=\f{e_r}{\sqrt{\Za}}\mu^{\ep/2}
\label{EQeren}
\eeqn
where, in the last equation, we have taken into account the relation
$\Ze=\Zphi$, which is a consequence of a Ward identity. Intuitively, it comes
from the
requirement $D_{\mu}\phi\rightarrow \sqrt{\Zphi}D_{\mu}\phi_r$ for the
covariant derivative. In the above equations, the bare quantities are on the
left-hand-side and the renormalized ones are on the right-hand-side, indicated
by the subscript ``r''.

We also note that the vacuum requires a special treatment \cite{ka96}.
For $\Phi=0$, the dimensionality requires $V\propto m^D$. We therefore
add a term $m^4h/g$ to the Lagrangian. With this we define a new
renormalization constant absorbing the vacuum divergencies by
\be
\f{m^4}{g}h&=&\f{m^4_r}{g_r\mu^{\ep}}\Zv
\ee

In the next Section, we calculate the different loop orders, based
on~(\ref{EQalgebraic}).

\section{Evaluation of the diagrams}
\label{calculation}

In the following, the subscript indicating the renormalized quantities will be
omitted, for brevity of notation. The different renormalization constants
are expanded with respect to the loop-parameter:
\be
Z_j=1+\sum_{l=1}^L
\hbar^l Z_j^{(l)}.
\label{EQrenconstexpansion}
\ee
The results are then identified order by oder in $\hbar$. To fix the ideas,
we give here to procedure for the gauge field trace-ln:
The bare diagram has to be modified according
to: $e^2\Phi^2\rightarrow e^2\Phi^2 \Zphi/\Za$, and the result reads, to
the first order in $\hbar$
\be
\Tr\ln(D^T)=\Tr\ln\left(-\p^2+e^2\Phi^2\f{\Zphi}{\Za}\right)
\approx\Tr\ln(-\p^2+e^2\Phi^2)+\hbar e^2\Phi^2(\Zphi^{(1)}-\Za^{(1)})\Tr
\left(
\f{1}{-\p^2+e^2\Phi^2}
\right).
\ee
The last term will contribute to the two-loop result, removing parts of its
$\ep$-poles.

\subsection{Renormalized zero- and one-loop order}

The renormalized zero-loop order is trivial

\be
V(l=0)=\f{1}{2}m^2\Phi^2+\f{g}{4!}\muep\Phi^4+\f{m^4}{g\muep},
\label{EQorder0}
\ee
while the renormalized one is simply a combination of the previous bare
zero-order and the trace-ln terms:
\beqn
V(l=1)&=&\f{1}{2}m^2\Phi^2\Zm^{(1)}+\f{g}{4!}\muep\Phi^4\Zg^{(1)}
+\f{m^4}{g\muep}\Zv^{(1)}\nonumber\\
&&\mbox{}+
\f{\Gamma(1-D/2)}{(4\pi)^{D/2}}\f{1}{D}\mu^{-\ep}
\left[
(N-1)(m_T^2)^2\left(\f{\mu^2}{m_T^2}\right)^{\ep/2}+
(m_L^2)^2\left(\f{\mu^2}{m_L^2}\right)^{\ep/2}+
(D-1)(\mg^2)^2\left(\f{\mu^2}{\mg^2}\right)^{\ep/2}
\right].
\label{EQorder1}
\eeqn
The constants $Z_j^{(1)}$ are determined in order to remove the $\ep$-poles
at $D=4$ in the Euler $\Gamma(1-D/2)$ function. In four dimensions, it would
be sufficient to make the $\ep$-expansion of~(\ref{EQorder1}) up to the order
$\ep^0$. However, we are also interested  in going to dimension $D=3$. For
this reason, it will be necessary to perform the $\ep$-expansion up to the
order $\ep$.
\cm{, according to the rule $l_{\ep}+l_{\hbar}=L$, where $L$ is the
maximum loop order and $l_{\ep}$ and $l_{\hbar}$ are the powers of $\ep$ and
$\hbar$ in the expansion.}

\subsection{Renormalized two-loop order}

The two-loop order contains contributions from the zero- and one-loop bare
diagrams, which will cancel the two-loop poles from the bare two-loop diagrams.
These contributions read (``ct'' is for counter-terms)
\be
V(l=2)^{\mbox{ct}}&=&
\f{1}{2}m^2\Phi^2\Zm^{(1)}+\f{g}{4!}\muep\Phi^4\Zg^{(1)}
+\f{m^4}{g\muep}\Zv^{(1)}\nonumber\\
&&\mbox{}+\f{\Gamma(1-D/2)}{(4\pi)^{D/2}}\f{1}{2}\mu^{-\ep}
\Biggl\{
(N-1)(m_T^2)\left(\f{\mu^2}{m_T^2}\right)^{\ep/2}
\left[
\f{1}{2}(3m_T^2-m_L^2)\Zm^{(1)}+\f{1}{2}(m_L^2-m_T^2)\Zg^{(1)}-m_T^2\Zphi^{(1)}
\right]\nonumber\\
&&\mbox{}+
(m_L^2)\left(\f{\mu^2}{m_L^2}\right)^{\ep/2}
\left[
\f{1}{2}(3m_T^2-m_L^2)\Zm^{(1)}+\f{3}{2}(m_L^2-m_T^2)\Zg^{(1)}-m_L^2\Zphi^{(1)}
\right]\nonumber\\
&&\mbox{}+
(D-1)(\mg^2)\left(\f{\mu^2}{\mg^2}\right)^{\ep/2}\left[
\mg^2(\Zphi^{(1)}-\Za^{(1)})\right]
\Biggr\}.
\label{EQct2loops}
\ee

To simplify the evaluation of the two-loop diagrams, we introduce the functions
$J(m)$ and $I(m_1,m_2,m_3)$, where we have kept
the same notation as in \cite{jones92}:
\beqn
J(m)&=&\int \dbarp \Delta(p,m),\\
I(m_1,m_2,m_3)&=& \int \dbarp \dbarq \Delta(p,m_1)\Delta(q,m_2)
\Delta(p+q,m_3).
\eeqn
The function $J(m)$ is trivial to determine, and its result
has in fact been used in~(\ref{EQct2loops}). For $I(m_1,m_2,m_3)$, we use
the result obtained in \cite{jones92},
which are in accordance with \cite{tausk92} and the recent
work \cite{davydychev00}.
The latter reference gives an expansion for all order in
$\ep$ but, since we need in the present work only the term of order $\ep^0$,
we stop at this order:
\beqn
J(m_1)&=&\f{\Gamma(1-D/2)}{(4\pi)^{D/2}}(m^2)^{D/2-1},\\
(\muep)^2(4\pi)^4I(m_1,m_2,m_3)&=&-\f{2}{\ep^2}(m_1^2+m_2^2+m_3^2)
-\f{2}{\ep}\left[\f{3}{2}(m_1^2+m_2^2+m_3^2)-L_1(m_1,m_2,m_3)
\right]\nonumber\\
&&\hspace{-3.5cm}\mbox{}-\f{1}{2}
\bigg\{L_2(m_1,m_2,m_3)-6L_1(m_1,m_2,m_3)+(m_2^2+m_3^2-m_1^2)
\barln(m_2^2)\barln(m_3^2)+(m_3^2+m_1^2-m_2^2)\barln(m_3^2)\barln(m_1^2)
\nonumber
\\
&&\hspace{-3.5cm}\mbox{}
+(m_1^2+m_2^2-m_3^2)\barln(m_1^2)\barln(m_2^2)+\xi(m_1,m_2,m_3)
+(m_1^2+m_2^2+m_3^2)[7+\zeta(2)]
\bigg\}
\label{EQIorder0}
\eeqn
where we have defined
\beqn
\barln(m^2)&=&\ln\left(\f{m^2}{\mu^2}\right)+\gamma-\ln(4\pi),\\
L_1(m_1,m_2,m_3)&=&m_1^2\barln(m_1^2)+m_2^2\barln(m_2^2)+m_3^2\barln(m_3^2),\\
L_2(m_1,m_2,m_3)&=&m_1^2[\barln(m_1^2)]^2+m_2^2[\barln(m_2^2)]^2+m_3^2
[\barln(m_3^2)]^2,\\
\xi(m_1,m_2,m_3)&=&4(2m_1^2m_2^2+2m_1^2m_3^2+2m_2^2m_3^2-m_1^4
-m_2^4-m_3^4)^{1/2}\left[L(\theta_1)+L(\theta_2)+L(\theta_3)
-\f{\pi}{2}\ln(2)\right].
\label{EQxilt0}
\eeqn
In the latter expression, $L(t)$ is the Lobachevsky function, defined
as
\be
L(t)=-\int_0^t dx \ln \cos x,
\label{EQlobalt0}
\ee
and the angles are given by
\be
\theta_j=\arctan\left[\f{(m_1^2+m_2^2+m_3^2)-2m_j^2}
{(2m_1^2m_2^2+2m_1^2m_3^2+2m_2^2m_3^2-m_1^4
-m_2^4-m_3^4)^{1/2}}\right].
\label{EQangleslt0}
\ee
These expressions, as well as the function $\xi(m_1,m_2,m_3)$, are valid
 for a positive
argument of the square root. This depends on the value of the masses. For
a negative value, one has to substitute
\beqn
\xi(m_1,m_2,m_3)&=&4(m_1^4+m_2^4+m_3^4-2m_1^2m_2^2
-2m_1^2m_3^2-2m_2^2m_3^2)^{1/2}\left[-M(-\theta_1)+M(\theta_2)+M(\theta_3)
\right],
\label{EQxigt0}
\\
M(t)&=&-\int_0^t dx \ln \sinh x,
\label{EQlobagt0}
\\
\theta_j&=&\coth^{-1}\left[\f{(m_1^2+m_2^2+m_3^2)-2m_j^2}
{(m_1^4+m_2^4+m_3^4-2m_1^2m_2^2-2m_1^2m_3^2-2m_2^2m_3^2)^{1/2}}\right].
\label{EQanglesgt0}
\eeqn
In the following, we shall always keep the symbolic notation
$\xi(m_1,m_2,m_3)$.
This will
avoid to check the sign of the argument of the square root.
It is not necessary to do so in the pure $\phi^4$-case where we know
that $m_L>m_T$, and, then, where the representation~(\ref{EQxilt0}) is valid.
 In the present case, because we do not known the
fixed points, it is not possible to specify the value of the photon mass
vs. the transverse or longitudinal mass.
We however mention that several
simplifications arise when two masses are equal, or when one, or two masses,
are
vanishing. For two equal masses, say $m_1=m_2$, the argument of the
square root becomes $m_3^4(1-4m_1^2/m_3^2)$.
This case arises when investigating
the O($N$)-theory. The ratio of the masses is the ratio $4m_T^2/m_L^2$ which is
always smaller than unity.
The relevant equations are then~(\ref{EQxilt0}), (\ref{EQlobalt0})
 and~(\ref{EQangleslt0}).
The case of vanishing masses is a little bit problematic for the extraction
of the power $\ep^0$ because of a cancellation of a diverging part of
$M(t)$ with a corresponding $\barln_j$ in~(\ref{EQIorder0}). For this reason,
we also give the following integrals:
\beqn
(\muep)^2I(m_1,0,0)&=&\f{1}{(4\pi)^D}
\f{\Gamma(2-D/2)\Gamma(3-D)\Gamma(D/2-1)^2}
{\Gamma(D/2)}m_1^2\left(
\f{m_1^2}{\mu^2}
\right)^{-\ep},
\label{EQm2m3null}\\
(\muep)^2I(m_1,m_1,0)&=&\f{1}{(4\pi)^D}\f{\Gamma(2-D/2)\Gamma(1-D/2)}
{D-3}m_1^2\left(
\f{m_1^2}{\mu^2}
\right)^{-\ep},
\label{EQm1m2gleichm3null}\\
(\muep)^2(4\pi)^4I(m_1,m_2,0)&=&-\f{2}{\ep^2}(m_1^2+m_2^2)-
\f{2}{\ep}\left[
\f{3}{2}(m_1^2+m_2^2)-L_1(m_1,m_2,0)
\right]\nonumber\\
&&\hspace{-3cm}\mbox{}-\f{1}{2}
\Biggl\{
L_2(m_1,m_2,0)-6L_1(m_1,m_2,0)+2m_1^2\barln(m_1^2)\barln(m_2^2)
+[\barln(m_1^2-m_2^2)]^2(m_1^2-m_2^2)\nonumber\\
&&\hspace{-3cm}\mbox{}-2\barln(m_1^2-m_2^2)\barln(m_2^2)(m_1^2-m_2^2)
+2(m_1^2-m_2^2)\mbox{Li}_2\left(
\f{m_2^2}{m_2^2-m_1^2}
\right)+(m_1^2+m_2^2)[7+\zeta(2)]+\f{\pi^2}{3}(m_1^2-m_2^2)
\Biggr\},
\label{EQIm1m2}
\eeqn
where $\mbox{Li}_2(z)=\sum_{i=1}^{\infty}z^i/i^2$ is the dilogarithm. It is
a simple exercice to check that, using $m_1=m_2$ in~(\ref{EQIm1m2}), we
recover the $\ep$-expansion of~(\ref{EQm1m2gleichm3null}), while using
 $m_3=0$, we recover the $\ep$-expansion of~(\ref{EQm2m3null}).
In the previous determination of~(\ref{EQIm1m2}), we have assumed that
$m_1>m_2$. The case $m_2>m_1$ is obtained from the former by the
replacement $m_1\leftrightarrow m_2$. This only affects the term of
order $\ep^0$. Comparing Eq.~(\ref{EQIm1m2}) with Eq.~(\ref{EQIorder0}), we
can also extract the way $\xi(m_1,m_2,m_3)$ diverges as one of the masses, say
$m_3$, goes to zero. For $m_1>m_2$, we have
\beqn
\lim_{m_3\rightarrow0}\left\{\xi(m_1,m_2,m_3)+(m_1^2-m_2^2)
[\barln(m_1^2)-\barln(m_2^2)]\barln(m_3^2)\right\}&=&\nonumber\\
&&\hspace{-8cm}(m_1^2-m_2^2)\left\{
\barln(m_1^2)\barln(m_2^2)+[\barln(m_1^2-m_2^2)]^2
-2\barln(m_1^2-m_2^2)\barln(m_2^2)+2\mbox{Li}_2\left(
\f{m_2^2}{m_2^2-m_1^2}
\right)+\f{\pi^2}{3}
\right\},
\label{EQxi1mass0}
\eeqn
while the opposite case  $m_1<m_2$ is obtained by permuting $m_1$ and $m_2$
in this equation.
%In the symmetric phase, for which $\Phi=0$, we shall need also to know
%$\xi(m_1,m_3,m_3)$ as $m_3\rightarrow0$. This can be readily obtained
%from~(\ref{EQxi1mass0}):
%
{}From (\ref{EQxi1mass0}) we can also determine  the case of two
simultaneously vanishing masses:
\be
\lim_{m_3\rightarrow0}\left\{\xi(m_1,m_3,m_3)+2m_1^2\barln(m_1^2)
\barln(m_3^2)-
m_1^2[\barln(m_3^2)]^2\right\}=m_1^2\left\{\barln(m_1^2)]^2+\f{\pi^2}{3}
\right\}.
\label{EQxi2mass0}
\ee
With all these definitions, we are now armed to compute the two-loop diagrams.
They are given by 
\begin{fmffile}{graph2}
\fmfset{wiggly_len}{2mm}
\fmfset{thin}{0.7pt}
\fmfset{dot_size}{1.0thick}
\beqn
\parbox{15mm}{\begin{center}
\begin{fmfgraph*}(35,25)
\fmfleft{i1}
\fmfright{o1}
\fmf{plain,left=1}{i1,v1,i1}
\fmf{plain,left=1}{o1,v1,o1}
\fmfdot{v1}
\end{fmfgraph*}\end{center}}&=&
\f{g}{4!}\muep\left\{
\left[
(N-1)J(m_T)+J(m_L)
\right]^2+2\left[
(N-1)J(m_T)^2+J(m_L)^2
\right]
\right\},
\\
\parbox{15mm}{\begin{center}
\begin{fmfgraph*}(35,25)
\fmfleft{i1}
\fmfright{o1}
\fmf{plain,left=1}{i1,v1,i1}
\fmf{photon,left=1}{o1,v1,o1}
\fmfdot{v1}
\end{fmfgraph*}\end{center}}
&=&\f{e^2}{2}\muep(D-1)J(\mg)
\left[
(N-1)J(m_T)+J(m_L)
\right],\\
\parbox{14mm}{\begin{center}
\begin{fmfgraph}(25,25)
\fmfleft{v1}
\fmfright{v2}
\fmf{plain}{v1,v2}
\fmf{plain,left=.7}{v2,v1}
\fmf{plain,left=.7}{v1,v2}
\fmfdot{v1,v2}
\end{fmfgraph}
\end{center}}
&=&\left(\f{g\muep}{3!}\right)^2\Phi^2
\left[
(N-1)I(m_T,m_T,m_L)+3I(m_L,m_L,m_L)
\right],
\\
\parbox{14mm}{\begin{center}
\begin{fmfgraph}(25,25)
\fmfleft{v1}
\fmfright{v2}
\fmf{plain}{v1,v2}
\fmf{photon,left=.7}{v2,v1}
\fmf{photon,left=.7}{v1,v2}
\fmfdot{v1,v2}
\end{fmfgraph}
\end{center}}
&=&4\left(\f{e^2\muep}{2}\right)^2\Phi^2
\Bigg\{
I(\mg,\mg,m_L)\left[
(D-2)+\f{1}{4\mg^4}(m_L^2-2\mg^2)^2
\right]\nonumber\\
&&
-\f{1}{4\mg^4}\left[
J(\mg)^2(m_L^2-2\mg^2)+2\mg^2J(\mg)J(m_L)+2(m_L^2-\mg^2)^2I(\mg,m_L,0)
-m_L^4I(m_L,0,0)
\right]
\Bigg\},\\
\parbox{14mm}{\begin{center}
\begin{fmfgraph}(25,25)
\fmfleft{v1}
\fmfright{v2}
\fmf{photon}{v1,v2}
\fmf{plain,left=.7}{v2,v1}
\fmf{plain,left=.7}{v1,v2}
\fmfdot{v1,v2}
\end{fmfgraph}
\end{center}}
&=&e^2\muep\f{1}{4}
\Bigg\{
(N-2)\left[
2J(\mg)J(m_T)-J(m_T)^2+I(\mg,m_T,m_T)(\mg^2-4m_T^2)
\right]\nonumber\\
&&+2
\bigg[
\f{(\mg^2+m_L^2-m_T^2)}{\mg^2}J(\mg)J(m_L)+
\f{(\mg^2-m_L^2+m_T^2)}{\mg^2}J(\mg)J(m_T)-J(m_L)J(m_T)\nonumber\\
&&-I(m_T,m_L,0)\f{(m_T^2-m_L^2)^2}{\mg^2}
+I(\mg,m_T,m_L)\f{(m_T^2-m_L^2)^2+\mg^2(\mg^2-2m_T^2-2m_L^2)}{\mg^2}
\bigg]
\Bigg\}
\eeqn
\end{fmffile}
where, compared to Eq.~(\ref{EQdiagrams}), we have included in the same 
diagram the different transverse and longitudinal components, and where the
vertices and corresponding multiplicities have been once again specified.

\section{Effective potential}
\label{effpot}

Collecting the results from the previous section,
the renormalization coupling constants are tuned to cancel the poles in
$1/\ep^2$ and $1/\ep$. Working with the effective potential, as we do here,
logarithms are appearing. Some of them have poles in $\ep$ coefficients.
These logarithms with pole coefficients have been named dangerous poles
by Chung and Chung \cite{chung97}. The cancellation of these dangerous poles
is a non trivial consequence
of the renormalizability of the theory.
We can see it as follow: Asking for the cancellation of one-loop poles,
we identify straightforwardly
\beqn
\Zm^{(1)}&=&
g\f{(N+2)}{3\ep},
\label{EQZm1}\\
g\Zg^{(1)}&=&
\f{g^2(N+8)+108e^4}{3\ep},
\label{EQZg1}\\
\Zv^{(1)}&=&
g\f{N}{2\ep},
\label{EQZv1}
\eeqn
where a factor $1/(4\pi)^2$ has been absorbed in the definition of $\hbar$
(see the expansion~(\ref{EQrenconstexpansion})).

The cancellation of the poles at the two-loop order gives renormalization
coefficients $\Zm^{(2)},\Zv^{(2)}$ which depend on
 $\barln(m_L),\barln(m_T)$ and $\Zphi^{(1)}$ and a renormalization coefficient
$\Zg^{(2)}$ which depends on $\barln(m_L),\barln(m_T),\barln(\mg)$ and
 $\Zphi^{(1)},\Za^{(1)}$.
We find that, asking for the independence with respect to these
dangerous logarithms,
fixes $\Zphi^{(1)},\Za^{(1)}$. This is non trivial since there are more
conditions than renormalization coefficients, and this is the mentioned
consequence of the renormalization. Everything together, the
cancellation of the poles at the two-loop order gives then
\beqn
\Zm^{(2)}&=&
\f{(N+2)}{9\ep^2}\left[g^2(N+5)-18ge^2+54e^4\right]
-\f{1}{6\ep}\left[g^2(N+2)-8ge^2(N+2)-6e^4(5N+1)\right]
\label{EQZm2}
\\
g\Zg^{(2)}&=&
\f{1}{9\ep^2}\left[
g^3(N+8)^2-18g^2e^2(N+8)+108g e^4(N+8)^2+108 e^6(N+18)
\right]
\nonumber\\
&&\mbox{}
-\f{1}{9\ep}\left[
g^3(5N+22)-12g^2e^2(N+5)-18ge^4(5N+13)+18e^6(7N+90)
\right]
\label{EQZg2}
,\\
\Zv^{(2)}&=&
g\f{N}{6\ep^2}[-18e^2+g(N+2)]+ge^2N\f{2}{\ep}.\\
\Zphi^{(1)}&=& e^2\f{6}{\ep},\\
\Za^{(1)}&=&-e^2\f{N}{3\ep},
\label{EQZv2}
\eeqn
where, as for the one-loop renormalization constants, a factor
$1/(4\pi)^2$ has also been absorbed in the definition of
$\hbar$.

With these one-loop and two-loop renormalization constants, all the poles
disappear from the $\ep$-expansion of the theory. Up to this order
(two loops), and,
in view of applications of this theory to phase transitions
in three dimensions,
the one-loop order  term has to be developed to the order $\ep$.
The two-loop effective potential in the Landau gauge can then be written as
\be
V=V^{(0)}+\f{\hbar}{(4\pi)^2} [V^{(1,0)}+\ep V^{(1,\ep)}]
+\left[\f{\hbar}{(4\pi)^2}\right]^2 V^{(2)},
\ee
where we have specified that a factor $(4\pi)^2$ is absorbed in the definition
of $\hbar$, and
with
\beqn
V^{(0)}&=&\f{1}{2}m^2\Phi^2+\f{g}{4!}\muep\Phi^4+\f{m^4}{g\muep},
\label{EQVorder0}\\
V^{(1,0)}&=&\f{\mu^{-\ep}}{8}\left\{
(N-1)m_T^4[-3+2\barln(m_T^2)]+m_L^4(-3+2\barln(m_L^2)]
+\mg^4[-5+6\barln(\mg^2)]
\right\},
\label{EQVorder1ep0}\\
V^{(1,\ep)}&=&-\f{\mu^{-\ep}}{96}
\Bigl(
(N-1)m_T^4\left\{21-18\barln(m_T^2)+6[\barln(m_T^2)]^2+\pi^2\right\}
+m_L^4\left\{21-18\barln(m_L^2)+6[\barln(m_L^2)]^2+\pi^2\right\}\nonumber\\
&&\mbox{}+3\mg^4\left\{9-10\barln(m_T^2)+6[\barln(m_T^2)]^2+\pi^2\right\}
\Bigr).\label{EQVorder1ep1}\\
\eeqn
The term $V^{(2)}$ is much longer to write. For this reason, we
give its expansion on the $\barln$-bases. With the definition
\be
V^{(2)}=\mu^{-\ep}\sum_{i=0}^2\sum_{j=0}^2\sum_{k=0}^2
V^{(2)}_{i,j,k}[\barln(m_T^2)]^i[\barln(m_L^2)]^j
[\barln(\mg^2)]^k,
\label{EQV2}
\ee
we have (see however the remark after these equations)
\beqn
V^{(2)}_{0,0,0}&=&
\f{1}{432g\mg^4}
\Biggl[
18\mg^2\biggl(
g^2\mg^2\left[
(N-1)(N+5)m_T^4+18m_L^2m_T^2-(2N+13)m_L^4
\right]\nonumber\\
&&\mbox{}+ge^2\{
42(m_L^2-m_T^2)^2(m_L^2+m_T^2)
-6\mg^2\left[
(19N-31)m_T^4+24m_T^2m_L^2+7m_L^4
\right]\nonumber\\
&&\mbox{}-18\mg^4\left[
(N-1)m_T^2+m_L^2
\right]+(19N-36)\mg^6
\}
-108e^4\mg^2(m_L^2-m_T^2)\left[
(N-1)m_T^2+3m_L^2
\right]
\biggr)\nonumber\\
&&\mbox{}+6g\Phi^2\left\{
7g^2\mg^4\left[
2(N-1)m_T^2+(N+8)m_L^2\right]+18e^4(7m_L^6-7m_L^4\mg^2+11m_L^2\mg^4+54\mg^6)
\right\}\nonumber\\
&&\mbox{}+\pi^2\biggl(
g^2\mg^4\left[
2(N-1)m_T^2+(N+8)m_L^2
\right]\left[
3(m_T^2-m_L^2)+g\Phi^2
\right]\nonumber\\
&&\mbox{}+18ge^2\mg^2\left\{
(m_L^2-m_T^2)^2(m_L^2+m_T^2)+3\mg^4\left[
(N-1)m_T^2+m_L^2
\right]-9\mg^6
\right\}\nonumber\\
&&\mbox{}+18e^4
\left\{
-9\mg^4(m_L^2-m_T^2)\left[
(N-1)m_T^2+3m_L^2
\right]+g\Phi^2(2m_L^6-m_L^4\mg^2+5m_L^2\mg^4+10\mg^6)
\right\}
\biggr)\nonumber\\
&&\mbox{}+216(4\pi)^4ge^2\left[
\mg^2(m_L^2-m_T^2)^2I^{(0)}(m_L,m_T,0)+e^2\Phi^2(\mg^2-m_L^2)^2
I^{(0)}(\mg,m_L,0)
\right]\nonumber\\
&&\mbox{}+54ge^2\biggl(
\mg^2\{
(N-2)(\mg^2-4m_T^2)\mg^2\xi(\mg,m_T,m_T)\nonumber\\
&&\mbox{}+2\left[
m_T^4-2m_T^2(\mg^2+m_L^2)+(\mg^2-m_L^2)^2
\right]\xi(\mg,m_T,m_L)
\}\nonumber\\
&&\mbox{}+e^2\Phi^2(m_L^4-4m_L^2\mg^2+12\mg^4)\xi(\mg,\mg,m_L)
\biggr)\nonumber\\
&&\mbox{}+6g^3\Phi^2\mg^4
\left[
(N-1)\xi(m_T,m_T,m_L)+3\xi(m_L,m_L,m_L)
\right]
\Biggr],
\label{EQV000}
\\
V^{(2)}_{1,0,0}&=&-\f{m_T^2}{12g\mg^2}
\biggl(
g^2(N-1)\mg^2[-m_L^2+(N+3)m_T^2+2g\Phi^2]
\nonumber\\
&&\mbox{}+6ge^2\left\{
3(m_L^2-m_T^2)^2-\mg^2[
2m_L^2+m_T^2(7N-9)]
+5(N-1)\mg^4
\right\}
-54e^4(N-1)\mg^2(m_L^2-m_T^2)
\biggr),
\label{EQV100}\\
V^{(2)}_{2,0,0}&=&\f{1}{72g\mg^2}
\biggl(
g^2(N-1)\mg^2\left\{-m_L^2(6m_T^2+g\Phi^2)
+m_T^2[3m_T^2(N+3)+4g\Phi^2]\right\}
\nonumber\\
&&\mbox{}-9ge^2\left[
-2(m_L^2-m_T^2)^2m_T^2+\mg^2(N-2)(\mg^4+6m_T^4)-6(2N-3)\mg^4m_T^2
\right]\nonumber\\
&&\mbox{}-162e^4(N-1)\mg^2(m_L^2-m_T^2)m_T^2
\biggr),
\label{EQV200}\\
V^{(2)}_{0,1,0}&=&-\f{m_L^2}{12g\mg^4}
\biggl(
g^2\mg^4[-(N+5)m_L^2+(2N+7)m_T^2+(N+8)g\Phi^2]
\nonumber\\
&&\mbox{}+6ge^2\mg^2\left\{
3(m_L^2-m_T^2)^2-\mg^2[
5m_L^2+2m_T^2]
+5\mg^4
\right\}\nonumber\\
&&\mbox{}-6e^4
\left\{
\mg^4\left[27(m_L^2-m_T^2)-16g\Phi^2\right]
+3g\Phi^2m_L^2(2\mg^2-m_L^2)
\right\}
\biggr),
\label{EQV010}\\
V^{(2)}_{0,2,0}&=&\f{m_L^2}{72g\mg^4}
\biggl(
g^2\mg^4[-3(N+5)m_L^2+3(N+8)m_T^2+(N+17)g\Phi^2]
\nonumber\\
&&\mbox{}+18ge^2\mg^2\left[
(m_L^2-m_T^2)^2
+3\mg^4
\right]\nonumber\\
&&\mbox{}-9e^4
\left\{
2\mg^4\left[27(m_L^2-m_T^2)-7g\Phi^2\right]
+g\Phi^2m_L^2(4\mg^2-3m_L^2)
\right\}
\biggr),
\label{EQV020}\\
V^{(2)}_{0,0,1}&=&\f{e^2}{6\mg^2}
\Bigl\{
-9e^2m_L^4\Phi^2
-3\mg^2\left[
m_L^4+m_T^4-2m_L^2(m_T^2+4e^2\Phi^2)
\right]\nonumber\\
&&\mbox{}+
12\mg^4\left[
m_L^2+(N-1)m_T^2-6e^2\Phi^2
\right]
-\mg^6(4N-9)
\Bigr\},\label{EQV001}\\
V^{(2)}_{0,0,2}&=&\f{e^2}{8\mg^4}
\left[
-18\mg^8
+e^2\Phi^2
\left(
-m_L^6+8m_L^4\mg^2-22m_L^2\mg^4+40\mg^6
\right)
\right],
\label{EQV002}\\
V^{(2)}_{1,1,0}&=&\f{1}{36\mg^2}
\Bigl\{
g(N-1)m_L^2\mg^2(3m_T^2+g\Phi^2)\nonumber\\
&&\mbox{}-9e^2\left[
-(m_L^2-m_T^2)^2(m_L^2+m_T^2)
+3\mg^2(m_L^4+m_T^4)-3\mg^4(m_L^2+m_T^2)+\mg^6\right]\Bigr\},
\label{EQV110}\\
V^{(2)}_{1,0,1}&=&\f{e^2}{4\mg^2}
\left[
-(m_L^2-m_T^2)^3+3\mg^2(m_L^4-m_T^4)-3\mg^4(m_L^2-m_T^2)+(N-1)\mg^6
\right],
\label{EQV101}\\
V^{(2)}_{0,1,1}&=&\f{e^2}{4\mg^4}
\Bigl\{
e^2m_L^6\Phi^2+\mg^2\left[
m_L^6-m_L^4(3m_T^2+4e^2\Phi^2)
+3m_L^2m_T^4-m_T^6
\right]
\nonumber\\
&&\mbox{}+\mg^4\left[m_L^2(-3m_L^2+14e^2\Phi^2)+3m_T^4\right]
+3\mg^6(m_L^2-m_T^2)+\mg^8
\Bigr\},
\label{EQV011}
\eeqn
where the function $I^{(0)}(m_1,m_2,0)$ denotes the $\ep\rightarrow0$
non-diverging piece of~(\ref{EQIm1m2}), without the coefficient $(\muep)^2$.
Some part of this function
may be put in the other terms of the expansion~(\ref{EQV2}), because
they lead, for example, to $\barln(m_1)\barln(m_2)$. We have not proceed in
this way because, in the scalar QED case, it is not yet known
if $\mg> m_T$. One has to study first the fixed points of the system.
At the critical point, one knows that, in the $4-\ep$ formalism with 
$\ep$-expansion,
$m\rightarrow0$. At the critical point, one would then need to compare
$e^2$ to $g/6$. If $\mg> m_T$, one is allowed to use~(\ref{EQIm1m2})
with $m_1=\mg,m_2=m_T$. In the opposite case $\mg< m_T$, one has to
use~(\ref{EQIm1m2})
with $m_2=\mg,m_1=m_T$.
We also note that $I^{(0)}(m_1,m_2,0)$ generate terms containing
$\barln(m_1^2-m_2^2)$. These terms can never be put on the basis
defined in~(\ref{EQV2}).

To see the coherence of our effective potential, we shall look at the
 $\phi^4$-limit, which can be reached using
$e^2\rightarrow 0$.
\cm{The other limit is scalar QED in the symmetric phase,
for which $\Phi=0$ implies
$\mg=0, m_T=m_L=m$.}

\subsection{$\phi^4$ limit}

In the limit $e^2\rightarrow0$, which gives the $\phi^4$ model,
the renormalized coupling constants and the effective potential are much
simplified. The renormalization constants can be readily read off
Eqs.(\ref{EQZm1})--(\ref{EQZv2}) with $e^2=0$. For the effective potential,
it is necessary to clean the results given in
(\ref{EQVorder0})--(\ref{EQVorder1ep1}) and (\ref{EQV000})--(\ref{EQV011}).
Taking the limit $e^2\rightarrow0$, we obtain
\beqn
V^{(0)}&=&\f{1}{2}m^2\Phi^2+\f{g}{4!}\muep\Phi^4+\f{m^4}{g\muep},
\label{EQVorder0phi4}\\
V^{(1,0)}&=&\f{\mu^{-\ep}}{8}\left\{
(N-1)m_T^4[-3+2\barln(m_T^2)]+m_L^4(-3+2\barln(m_L^2)]
\right\},
\label{EQVorder1ep0phi4}\\
V^{(1,\ep)}&=&-\f{\mu^{-\ep}}{96}
\Bigl(
(N-1)m_T^4\left\{21-18\barln(m_T^2)+6[\barln(m_T^2)]^2+\pi^2\right\}
+m_L^4\left\{21-18\barln(m_L^2)+6[\barln(m_L^2)]^2+\pi^2\right\}
\Bigr).\label{EQVorder1ep1phi4}\\
\eeqn
The term $V^{(2)}$ is considerably simplified compared to the
full QED model. It has the expansion
\be
V^{(2)}=\mu^{-\ep}\sum_{i=0}^2\sum_{j=0}^2
V^{(2)}_{i,j}[\barln(m_T^2)]^i[\barln(m_L^2)]^j,
\label{EQV2phi4}
\ee
with the following coefficients:
\beqn
V^{(2)}_{0,0}&=&\f{g}{432}
\Bigl(
6\left\{
54m_L^2m_T^2-3(2N+13)m_L^4+7(N+8)m_L^2g\Phi^2
+(N-1)m_T^2\left[
3(N+5)m_T^2+14g\Phi^2
\right]
\right\}\nonumber\\
&&\mbox{}+\left[
2(N-1)m_T^2+(N+8)m_L^2
\right]\left[
3(m_T^2-m_L^2)+g\Phi^2
\right]\pi^2+6g\Phi^2\left[
(N-1)\xi(m_T,m_T,m_L)+3\xi(m_L,m_L,m_L)
\right]
\Bigr)\nonumber\\
&=&\f{g}{24}\bigl\{
m_T^4(N-1)(N-9)+m_L^4(5N+43)+m_L^2m_T^2(7N-52)\nonumber\\
&&\mbox{}+(m_L^2-m_T^2)\left[
(N-1)\xi(m_T,m_T,m_L)+3\xi(m_L,m_L,m_L)
\right]
\bigr\}
\label{EQV00phi4}
\\
V^{(2)}_{1,0}&=&-\f{g}{12}(N-1)m_T^2\left[
-m_L^2+(N+3)m_T^2+2g\Phi^2
\right]
\nonumber\\
&=&-\f{g}{12}(N-1)m_T^2\left[
5m_L^2+(N-3)m_T^2
\right]
\\
V^{(2)}_{2,0}&=&-\f{g}{72}(N-1)
\left\{
m_L^2(6m_T^2+g\Phi^2)-m_T^2[3(N+3)m_T^2+4g\Phi^2]
\right\}
\nonumber\\
&=&-\f{g}{24}(N-1)
\left[
m_L^4-3m_L^2m_T^2-(N-1)m_T^4
\right]
\\
V^{(2)}_{0,1}&=&-\f{g}{12}m_L^2\left[
-(N+5)m_L^2+(2N+7)m_T^2+(N+8)g\Phi^2
\right]\nonumber\\
&=&\f{g}{12}m_L^2\left[
(N+17)m_T^2-(2N+19)m_L^2
\right]
\\
V^{(2)}_{0,2}&=&\f{g}{72}
m_L^2\left[
-3(N+5)m_L^2+3(N+8)m_T^2+(N+17)g\Phi^2
\right]\nonumber\\
&=&\f{g}{8}m_L^2(4m_L^2-3m_T^2)
\\
V^{(2)}_{1,1}&=&\f{g}{36}(N-1)m_L^2(3m_T^2+g\Phi^2)\nonumber\\
&=&\f{g}{12}(N-1)m_L^4,
\label{EQV11phi4}
\\
\eeqn
where, in the second equality, we have replaced $g\Phi^2$ by $3(m_L^2-m_T^2)$.
We have checked that Eqs.~(\ref{EQV00phi4})--(\ref{EQV11phi4}) reproduce
the effective  potential given in \cite{jones92}.
The correct limit of the pure $\phi^4$-theory is then recovered.

\subsubsection*{Remark}

In the previous subsection, we have obtained the effective potential of
$\phi^4$-theory by taking the limit $e^2\rightarrow 0$ for fixed $\mg^2$, then
taking the limit $\mg^2\rightarrow 0$. This is the fastest way to obtain the
correct limit, but it is quite {\em cavalier} to proceed so.
We have  checked, replacing everywhere $\mg^2$ by its value
$\mg^2=e^2\Phi^2$, that it was justified to do so: all the $\mg^2$-terms in the
denominator of Eqs.~(\ref{EQV000})--(\ref{EQV011}) are properly
compensated to give the two-loop result
of Eqs~(\ref{EQV00phi4})--(\ref{EQV11phi4}).

%Rem: comme il n'y a pas de termes en $1/e^2$, il n'y a pas de dev. de Taylor
%a realiser, puisque il n'y a pas 1/e^2(f(0)+f'(0)e^2+f''(0)e^4+...):
%la fonction f(e^2) est identiquement nulle. Donc f(0) et toutes les derivees
%sont nulles. Dans le cas contraire, on aurait du garder dans les
%termes d'ordre (e^2)^0 la partie f'(0) puisque
%cela compense 1/e^2. C'est important pour la section suivante.
%Attention qu'il faut un developpement de Taylor ...

\section{Conclusion}
\label{conclusion}

In this paper, we have determined the effective potential of scalar QED in
$4-\ep$ dimensions in an $\ep$-expansion.
For zero charge, we revover the well-known result
for the pure O($N$)-symmetric $\phi^4$-theory.
The full effective potential will
be used in order to determine various amplitude ratios in the context
of a scalar field in the presence of a gauge field. This comprises
the superconducting case, for which $N=2$.
In this respect,
it might be interesting
to work in an arbitrary gauge and to show
that these amplitude ratios are gauge independent.
This can be done
as in Ref.~\cite{kang74}.

We shall also, in a subsequent paper, recast our result in the scaling form of
Widom which shows the critical behavior of the free energy explicitely.

An other interesting work
will be the application
of
 variational perturbation theory
to derive the various critical exponents in scalar QED, including the
amplitude ratios
along the lines of Ref. \cite{kleinertvdb01}.
In $4-\ep$ dimensions with $\ep$-expansion, this has not even been done
for the simpler O($N$)-symmetric $\phi^4$-model
without gauge field.

\begin{acknowledgments}
We are grateful to Dr. F. S. Nogueira for several
interesting discussions and clarifications.
We thank A. Pelster for several improvements of an earlier version
of the manuscript.
The work of B. VdB was supported by the
Alexander von Humboldt foundation and
the Institut Interuniversitaire des Sciences Nucl\'eaires de Belgique.
\end{acknowledgments}

\newpage

%\twocolumn

%\input fig.tex


\begin{thebibliography}{99}
\bibitem{jones92}
C. Ford, I. Jack, and D. R. T. Jones, Nucl. Phys. B {\bf 387}, 373 (1992).
\bibitem{tausk92}
A. I. Davydychev, and J. B. Tausk, Nucl. Phys. B {\bf 397}, 123 (1993).
\bibitem{davydychev00}
A. I. Davydychev, Phys. Rev. D {\bf 61}, 087701 (2000).
\bibitem{tri}
 H. Kleinert,
     Lett.\ Nuovo Cimento   {\bf 35}, 405 (1982)
(http://www.physik.fu-berlin.de/\~{}kleinert/97);\\
 H. Kleinert,
    {\em Gauge Fields in Condensed Matter\/},
    Vol.\ I \,\,  Superflow and Vortex Lines,
    World Scientific Publishing Co., Singapore 1989, pp. 735-742
(http://www.physik.fu-berlin.de/\~{}kleinert/b1).
%\bibitem{olsson98}
%P. Olsson and S. Teitel, Phys. Rev. Lett. {\bf 80}, 1964 (1998).
\bibitem{barto83}
J. Bartholomew, Phys. Rev. B {\bf 28}, 5378 (1983);
Y. Muneshisa, Phys. Lett. B {\bf 155}, 159 (1985).
\bibitem{halperin74}
B. I. Halperin, T. C. Lubensky, and S.-K. Ma, Phys. Rev. Lett. {\bf 32}, 292
(1974);\\ 
J.-H. Chen, T. C. Lubensky, and D. R. Nelson, Phys. Rev. B {\bf 17},
4274 (1978).
\bibitem{lawrie82}
I. A. Lawrie, Nucl. Phys. B {\bf 200}, 1 (1982).
\bibitem{tessman84}
J.-P. Tessman, {\em Two-loop Renormierung der skalaren Elektrodynamik},
Diplomarbeit, Freie Universit\"at Berlin,
written under the supervision of
one of the authors (H.K.)
\bibitem{folk96}
R. Folk, and Yu. Holovatch, J. Phys. A {\em 29}, 3409 (1996);\\
R. Folk, and Yu. Holovatch,
{\em Critical fluctuations in normal-to-superconducting transition},
Lecture at the 1st Winter Workshop "Cooperative Phenomena in Condensed Matter",
March, 1998, Pamporovo, Bulgaria ; cond-mat/9807421.
\bibitem{nog99}
M. Kiometzis, H. Kleinert, and A. M. J. Schakel, Phys. Rev. {\bf 73},
1975 (1994);\\
I. F. Herbut, and Z. Tesanovi{\'c},
Phys. Rev. Lett. {\bf 76}, 4588 (1996);\\
C. de Calan, A. P. C. Malbuisson, F. S. Nogueira, and N. F. Svaiter,
Phys. Rev. B {\bf 59}, 554 (1999);\\
C. de Calan, and F. S. Nogueira, Phys. Rev. B {\bf 60}, 4255 (1999) ;
B. Bergerhoff, F. Freire, D. F. Litim, S. Lola, and C. Wetterich,
Phys. Rev. B {\bf 53}, 5734 (1996).
\bibitem{kleinertpi}
H. Kleinert, {\em Path Integrals in Quantum Mechanics, Statistics
and Polymer Physics}, World Scientific. Variational perturbation
theory is developed in chapters 5 and 17.
\bibitem{kleinert257}
H. Kleinert, Phys. Rev. D {\bf 57}, 2264 (1998) ; Addendum  {\bf
  58} 107702 (1998).
\bibitem{kleinert279}
H. Kleinert, Phys. Rev. D {\bf 60}, 085001 (1999).
\bibitem{kleinert287}
F. Jasch, and H. Kleinert, J. Math. Phys.  {\bf 42}, 52 (2001).
\bibitem{kleinert263}
H. Kleinert, Phys. Lett. B {\bf 434}, 74 (1998);  {\bf 463}, 69
(1999).
\bibitem{kleinert295}
H. Kleinert, and V. Schulte-Frohlinde, J. Phys. A {\bf 34}, 1037 (2001).
\bibitem{kleinertvdb01}
H. Kleinert, and B. Van den Bossche, to be published in Phys. Rev. E;
{\em Three-loop critical exponents, amplitude functions and amplitude ratios
 from variational perturbation theory}, cond-mat/0011329
\bibitem{dohm99}
M. Str\"osser, L. A. Larin, and V. Dohm, Nucl. Phys. B {\bf 540},
654 (1999).
\bibitem{folk92}
S. Kolnberger, and R. Folk, Phys. Rev. B {\bf 41}, 4083 (1992).
\bibitem{brezin72}
E. Br\'ezin, D. J. Wallace, and K. G. Wilson, Phys. Rev. Let.
{\bf 29}, 591 (1972); Phys. Rev. B
{\bf 7}, 232 (1973).
\bibitem{zj76}
E. Br{\'ezin}, J. C. Le Guillou, and J. Zinn-Justin, in {\em Phase
transition and critical phenomana}, edited by C. Domb and M. S.
Green (Academic Press, New-York, 1976), Vol. 6., p. 125.
\bibitem{zia73}
D. J. Wallace, and R. K. P. Zia, Phys. Lett. A {\bf 46}, 261
(1973); J. Phys. C {\bf 7}, 3480 (1974).
\bibitem{DeWitt}
B. De Witt, Phys. Rev. {\bf 162}, 1195 (1967).
\bibitem{ka00}
B. Kastening, Phys. Rev. E {\bf 61}, 3501 (2000).
\bibitem{ka00b}
M. Bachman, A. Kleinert, and A. Pelster, Phys. Rev. D {\bf 61}, 085017 (2000);\\
H. Kleinert, A. Pelster, B. Kastening, and M. Bachman, Phys. Rev. E {\bf 62},
1537 (2000).
\bibitem{vdb01b}
H. Kleinert, A. Pelster, and B. Van den Bossche, in preparation.
\bibitem{ka96}
B. Kastening, Phys. Rev. D {\bf 54}, 3965 (1996);\\
B. Kastening, Phys. Rev. D {\bf 57}, 3567 (1998).
\bibitem{chung97}
J. M. Chung, and B. K. Chung, Phys. Rev. D {\bf 56}, 6508 (1997);\\
J. M. Chung, and B. K. Chung, J. Korean Phys. Soc. {\bf 33}, 643 (1998).
\bibitem{kang74}
J. S. Kang, Phys. Rev. D {\bf 10}, 3455 (1974).
\end{thebibliography}
\end{document}